\documentclass[fleqn,usenatbib]{mnras}
\usepackage{hyperref}
\hypersetup{draft}
\usepackage{color}
\usepackage{graphicx}
\usepackage{natbib}

\usepackage{multirow}   
\usepackage{xspace}
\usepackage{longtable}

\newcommand{\gaia}{{\it Gaia}}
\newcommand{\unit}[1]{\ensuremath{\mathrm{\,#1}}\xspace}
\newcommand{\masyr}{\unit{mas\,yr^{-1}}}

\title[Kinematics beats dust]{Kinematics beats dust: unveiling nested substructure in the perturbed outer disc of the Milky Way.}

\author[Laporte, Koposov \& Belokurov]{
\parbox[t]{\textwidth}{ Chervin F. P. Laporte$^{1}$, Sergey E. Koposov$^{2,3}$, Vasily Belokurov$^{3}$}
\\
$^{1}$ Kavli Institute for the Physics and Mathematics of the Universe (WPI),\\ The University of Tokyo Institutes for Advanced Study
(UTIAS), The University of Tokyo, Chiba 277-8583, Japan \\
$^{2}$ Institute for Astronomy, University of Edinburgh, Royal Observatory, Blackford Hill, Edinburgh EH9 3HJ, UK\\
$^{3}$Institute of Astronomy, University of Cambridge, Madingley road, CB3 0HA, UK\\
}
\date{Accepted . Received ; in original form }
\pubyear{2018}
\begin{document}
\label{firstpage}
\pagerange{\pageref{firstpage}--\pageref{lastpage}}
\maketitle
\begin{abstract}
We use the Gaia eDR3 data and legacy spectroscopic surveys to map the Milky Way disc substructure towards the Galactic Anticenter at heliocentric distances $d\geq10\,\rm{kpc}$. We report the discovery of multiple previously undetected new filaments embedded in the outer disc in highly extincted regions. Stars in these over-densities have distance gradients expected for disc material and move on disc-like orbits with $v_{\phi}\sim170-230\,\rm{km\,s^{-1}}$, showing small spreads in energy. Such a morphology argues against a quiescently growing Galactic thin disc. Some of these structures are interpreted as excited outer disc material, kicked up by satellite impacts and currently undergoing phase-mixing (``feathers''). Due to the long timescale in the outer disc regions, these structures can stay coherent in configuration space over several Gyrs. We nevertheless note that some of these structures could also be folds in the perturbed disc seen in projection from the Sun's location. A full 6D phase-space characterization and age dating of these structure should help distinguish between the two possible morphologies. 

\end{abstract}
\section{Introduction}

Large photometric surveys such as 2MASS, the Sloan Digital Sky Survey (SDSS) and PanSTARRS have revealed much tumult at the disc-halo interface through the discovery of numerous streams and cloud-like structures about the Galactic mid-plane out to large latitudes ($b<40^{\circ}$) \citep{newberg02,slater14}. Amongst these tributaries we note the Anticenter Stream and Eastern Banded Structure \citep[ACS, EBS][]{grillmair11}, Triangulum-Andromeda \citep[TriAnd][]{rocha-pinto04} and the Monoceros Ring \citep{newberg02,slater14}. These have originally been interpreted as ancient accretion events from shredded satellite galaxies \citep{penarrubia05} although a tidally excited disc origin has also been discussed at length \citep[e.g.][]{kazantzidis08}. The latter interpretation has now been confirmed thanks to new observational studies of stellar populations, kinematics and chemistry \citep[e.g.][]{price-whelan15, deboer17, bergemann18} combined with numerical modelling of the response of the Galactic disc to satellite galaxies \citep{gomez16,laporte18b}. While most of these structures have been identified far from the mid-plane, our knowledge of the low-latitude Galactic disc's structure through photometric surveys has been limited by the copious amounts of dust extinction. A way around this is to use kinematic information. A mission like \gaia\ is thus perfectly suited to deliver accurate enough proper motions in order to trace Galactic structure (and substructure) about the mid-plane out to large distances. This is the aim of this study.

In this Letter, we apply a simple phase-space mapping method to study the structure of the outer disc beyond $d>10$ kpc complementing the recent discoveries of moving groups in the solar neighbourhood \citep[e.g.][]{antoja18} and towards the Anticenter \citep{2021arXiv210105811G}. In Section 2, we our algorithm to identify prominent peaks in the proper motion space borrowed from methods used in cosmological simulations. In section 3, we validate the method by identifying known structures at $|b|>10^{\circ}$ such as prominent streams, outer disc substructures, globular clusters and dwarfs as well as the full extent of the Sgr trailing/leading arms. We focus on the Anticenter region in the midplane where extinction is much higher and reveal a number of striking filamentary and previously unknown structures in the Anticenter. We follow one of the newly uncovered structures as a case study (leaving a deeper investigation into all other structures to a future study), determining its distance variation using red clump (RC) and sub-giant branch tracers, metallicity and kinematic properties through complementary information from legacy surveys. In Section 4, we interpret our results through recent results from numerical N-body models of host satellite interactions and conclude in Section 5.

\begin{figure*}
\includegraphics[width=1.0\textwidth,trim=0mm 0mm 30mm 0mm, clip]{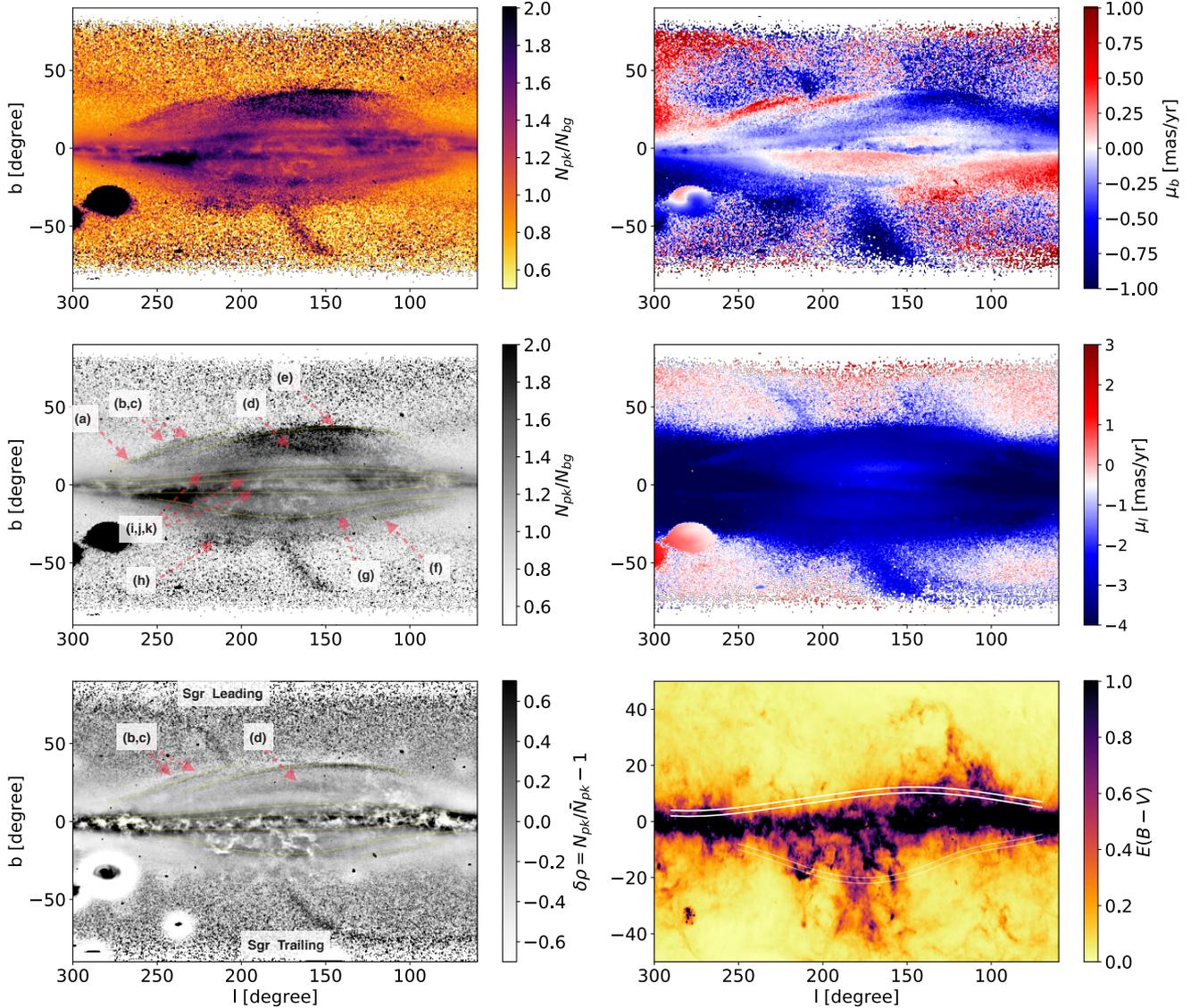}
\caption{{\it Top Left}: number of stars per square degree normalised by the background count. Three new prominent filamentary structures about the midplane are markedly visible. The algorithm also picks up a number a large number of globular clusters, satellites galaxies (classical dwarfs) and the Sgr stream as well as its bifurcation in kinematic space (see middle right panel). 
 {\it Middle Left} The various annotated features correspond to (a) Eastern Banded Structure (EBS), (b,c) Split about the EBS (see Grillmair 2011), (d) ``Monoceros Stream'' \citep[see][]{slater14,ramos21}, (e) Anticenter Stream, (f) LKB-01 ``Latifah'', (g) LKB-02 ``Ndegeocello'', (h) diffuse extension (i) LKB-03 ``Dumile'' (j) LKB-04 ``Shakur'' (k) LKB-05 `` Dilla''. {\it Bottom Left:} Unsharp masked image of the Galactic Anticenter. {\it Top Right}: Reflex motion corrected proper motion latitudinal proper motion $\mu_{b}(l,b)$. We apply a uniform correction assuming $d\sim13\,\rm{kpc}$ (this leaves some residuals in the halo). {\it Middle Right}: Reflex motion corrected longitudinal proper motion $\mu_{l}(l,b)$. We are also able to retrieve kinematic signal from the bifurcation of the Sgr stream towards $b<-40$, as well as a small portion at $b\sim60$ and $l\sim200$.  {\it Bottom Right}: Extinction map with splines delineating structures ``i'' and ``g''. The structures pass through regions of appreciably high extinction $E(B-V)>1$. This makes the identification of a sub-giant branch difficult outside the range $150<l<260$.} 
\end{figure*}

\section{Method:}

We start our search for outer disc substructure by isolating stars with $\varpi/\sigma_{\varpi}<6$, to select a sample of objects with low parallax measurements. We then keep stars with $\varpi-\sigma_{\varpi}<0.1\,\rm{mas}$, thus preferentially selecting objects beyond $d\sim10$ kpc. We grid the data in $(l,b)$ pixels of equal width of $\Delta=0.5^{\circ}\times0.5^{\circ}$ and identify prominent kinematic peaks in each pixel. This is done using the shrinking sphere algorithm \citep{Power2003} commonly used in analysing centers of bound structures in N-body simulations which we apply on 2D proper motion data $(\mu_{\alpha},\mu_{\delta})$ in each spatial pixel. This robustly picks out the most prominent proper motion peaks. Although in this work we reveal many new features, we note that it should also be possible to further track structures which spatially overlap by tracing multiple proper motion peaks (this will be the subject of a future contribution). For each detection, we select stars falling within a radius of $r_{\mu}<1.0$\,\masyr of the peak down to a magnitude limit of $G=21$. We also estimate the background density in proper motion space using a cylindrical annulus of width $\Delta R=0.6$\,\masyr centered around each peak ($r_{\mu}<1.0$\,\masyr).

\section{Results}
\subsection{Anticenter map \& new substructures}

\begin{figure}
\includegraphics[width=0.5\textwidth,trim=10mm 40mm 00mm 30mm, clip]{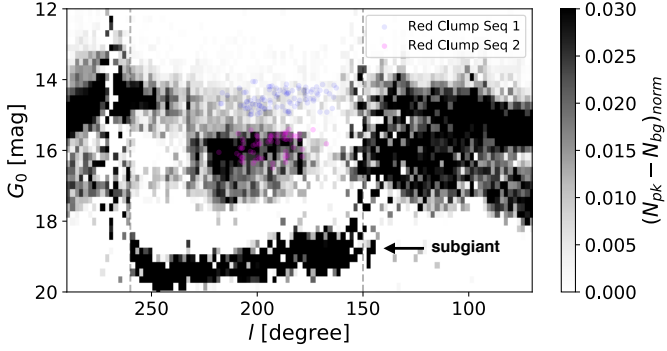}
\caption{Residual ($N_{pk}(G_0, l) - N_{bg}(G_0,l)$) of the column normalised histogram of the extinction corrected $G$-band magnitude for RC candidates as a function Galactic longitude $l$ along the structure ``i'' (LKB-03). The stars were selected to be within a spline tracing LKB-03 of width $\Delta b\sim2^{\circ}$, have colours consistent with RC ($0.85<(BP-RP)_0<1.2$) and $E(B-V)<1$. One notices two RC sequences, a bright and a faint one. The subtraction of the background signal also reveals a lower RGB sequence with a weak distance gradient. A number spectroscopic RC are identified in the LAMOST DR6 data for in the brighter and faint RC sequences, shown as the blue and magenta points respectively. Dashed grey lines delineate regions that are heavily affected by dust extinction (see also bottom right panel of Fig. 1).} 
\end{figure}

\begin{figure}
\includegraphics[width=0.5\textwidth,  clip]{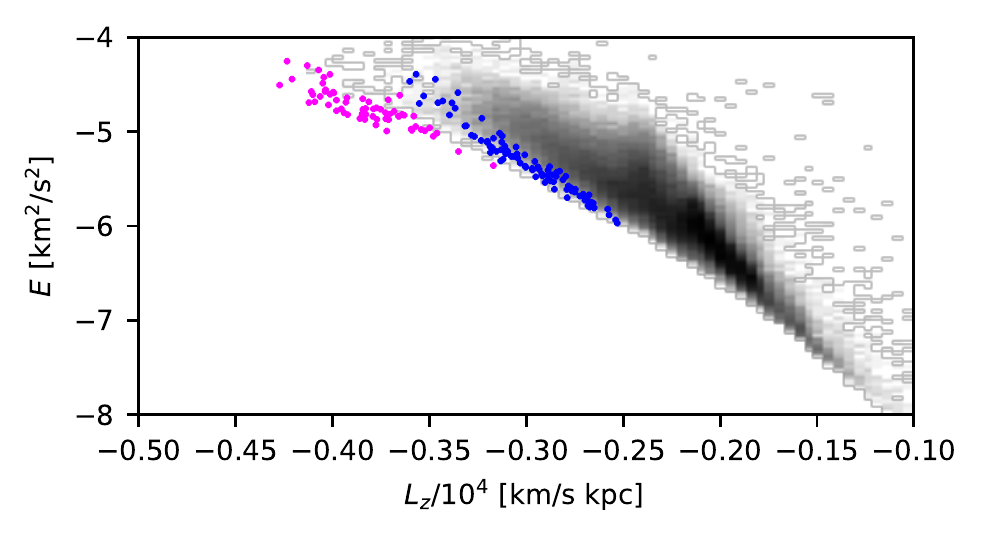}
\caption{Energy vs angular momentum for stars cross-matched between LAMOST DR6 with the \gaia\ catalog (greyscale histogram) and RC sequences in the structure ``i''. The fainter sequence (magenta points) has a smaller spread in energies compared to the brighter one (blue points).}
\end{figure}

\begin{figure}
\includegraphics[width=0.5\textwidth,  clip]{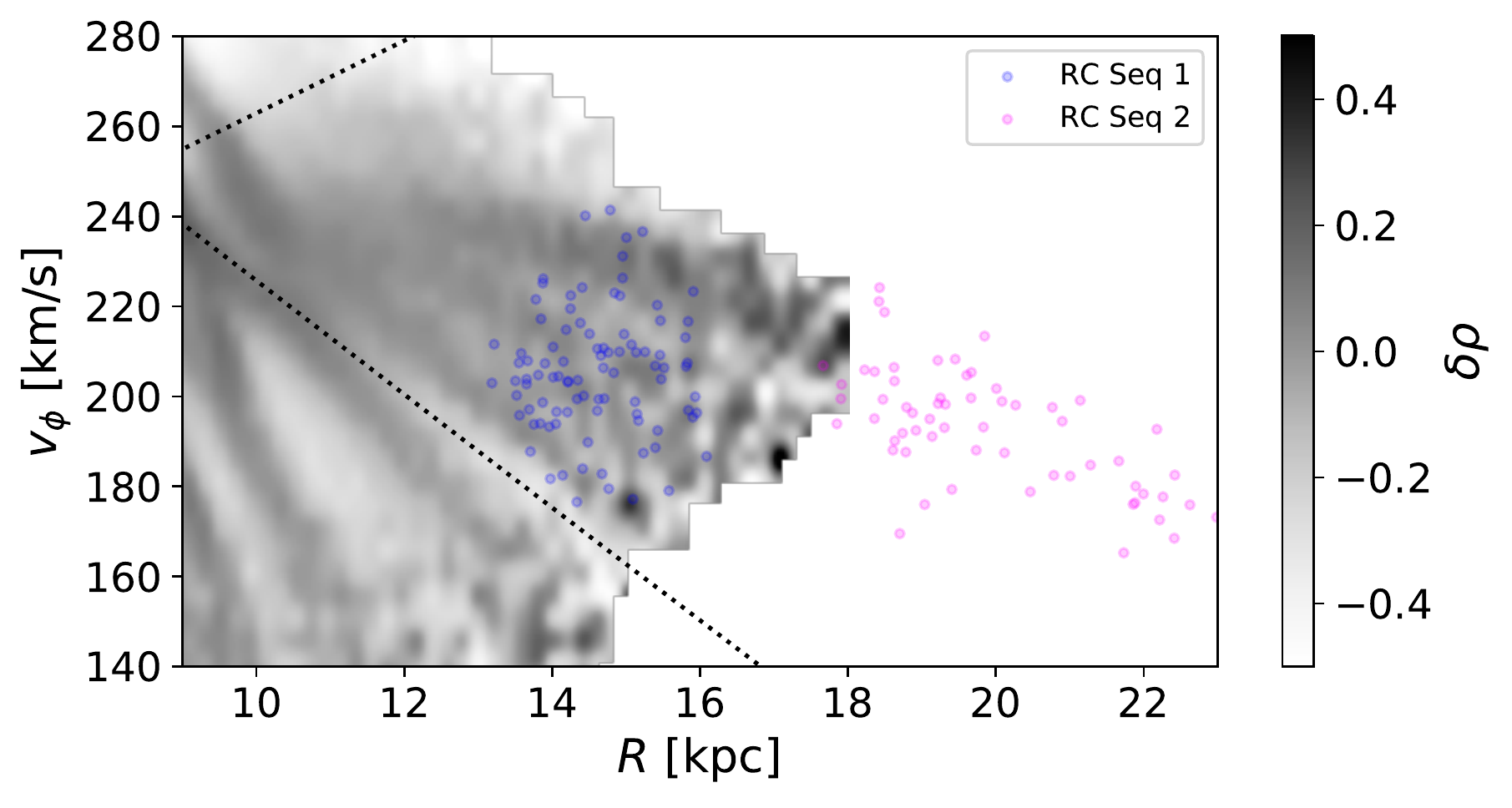}
\caption{$V_{\phi}$ vs R plot for a portion of the Anticenter as defined by \citep{2021arXiv210105811G}. We note the strong presence of the AC ridges 1,2. The bright RC sequence shows a large spread in velocities, whereas fainter sequence seems be an extension of their ``AC new ridge 2''. Overlaid as dotted lines are the lines assuming denoting our kinematic selection cone assuming a line of sight velocity of $v_{los}=0\, \rm{km\,s^{-1}}$ towards the Anticenter.}
\end{figure}

We apply our method to the field centered around the Galactic Anticenter which we present in Figure~1. The top left panel shows the relative contrast of the proper motion peaks in every pixel over the background. One notices immediately a striking threaded pattern buried in the mid-plane of the disc, made up of long filamentary structures across the mid-plane. This new and rich pattern is superposed with known asymmetric arching structures above and below the mid-plane reconnecting to the disc at much sharper density contrast. All new structures are labelled in Figure 1 in the middle left panel. The number of new structures is significant, not only {\it doubling} the current count of known thin substructures (e.g. ACS, EBS), but also shedding further light on the morphology of previously known ``overdensities''. We note that in the region of TriAnd \citep[e.g.][]{rocha-pinto04} there are now two distinct and extended filamentary structures (LKB-01, LKB-02) with the top one spanning from $l\sim70$ to $l\sim250$. We discuss its association with TriAnd in section 3.2. 

The bottom left panel of  Figure~1 shows an unsharp masked image of the Anticenter as $\delta=N_{pk}/<{N_{pk}}>-1$, where $<N_{pk}>$ is the image matrix smoothed with a Gaussian kernel of 8 pixels (the pixel size is $(0.5^{\circ})^2$). The top and middle right panels show the corresponding median proper motion maps of stars within each peak $\mu_l(l,b)$ and $\mu_b(l,b)$ for which we apply a solar reflex motion correction assuming a median distance of $d\sim13\,\rm{kpc}$. The proper motion maps show that the structures move on disc-like orbits ($\mu_l$) and that they oscillate ($\mu_b$); this is mostly appreciated for structures such as the EBS, ACS and structures LKB-01, LKB-02 where the amplitudes of oscillations are naturally largest. In particular, we note that $\mu_{b}$ traces the uncovered disc over-densities rather well and separates them from the background. For example, the split in EBS is clearly visible in $\mu_b$ at $(l,b)=(230, 28)$ and similarly for a number of the newly discovered substructures.

\subsection{Nested disc substructures}

To map each substructure presented in the previous section we construct $2^{\circ}$ wide splines tracing them on the sky.  We then analyse the magnitude distribution as a function of galactic latitude for stars selected to lie within the on-sky spline mask and with colours consistent with the RC $0.85<(\rm{BP-RP})_{0}<1.2$. We apply a cut of $E(B-V)<1$ to avoid highly extincted regions. For this Letter, we will focus only on one substructure (``i'', LKB-03 in Fig. 1) and defer a more detailed analysis on all structures to a future contribution (taking into account the presence of multiple kinematic peaks in each pixel). In Figure 2, we show the column normalised variation of the extinction corrected G-band magnitude $G_{0}$ as a function of longitude $l$ for structure ``i'' (LKB-03) using RC-like stars identified in peaks corrected for the background, shown as a grey-scale 2D histogram. We note the presence of two RC sequences, a bright one with $G_0\sim14.5$ and a fainter one with $G_0\sim15.7$ at $l=180$, corresponding to distances of $d\sim6\,\rm{kpc}$ and $d\sim11\rm{kpc}$ \citep{hawkins17}. In regions where the extinction is low ($150<l<250$, see bottom right panel of Fig 1), we able to pick out a sub-giant branch \citep[see also][]{belokurov06}, which also reveals a shallow distance gradient along the structure. Outside this region, we are heavily affected by the Galactic extinction hence why we lose the sub-giant. A common geometric behaviour shared by both RC sequences is the average trend where they are systematically closer to the observer in the Anticenter direction and move further away at their extremities as $l\to 70$ and $l\to300$. This behaviour is naturally expected when viewing a ring (or disc material) around the Galaxy from the Sun's perspective.

We next cross-match the stars in each structures to the LAMOST survey \citep{luo12} DR6 low resolution spectra catalog of A,F,G,K-type stars. Using the following cuts on our candidate photometrically selected RC ($0.85<(BP-RP)_{0}<1.2$):
\begin{itemize}
    \item  $2.2<\rm{log}(g)<2.75$
    \item $4600<T_{\rm{eff}}<5000$,
\end{itemize} we are able to select RCs \citep{huang20} in both the bright and the faint sequences in structure "i", shown in Figure 2 as the transparent blue and magenta points respectively. The stars have median metallicities typical of the outer disc \citep[e.g.][]{bergemann18,laporte20} with  $[\rm{Fe/H}]\sim-0.5$ and $[\rm{Fe/H}]\sim-0.6$ for the bright and faint RC sequences.

\begin{table}
\centering 
\begin{tabular}{c rrrrrrr} 
\hline\hline 
Structure& $l_1$ & $b_1$ & $l_2$ & $b_2$ & $l_3$ & $b_3$  \\
 & [deg] & [deg] & [deg] & [deg] & [deg] & [deg]\\
\hline 
LKB-01 & 80 & $-13.0$ & 100& $-17.0$& 120& $-21.0$\\ 
LKB-02 & 70 & $-5.6$ & 170& $-21.0$& 250& $-9$\\ 
LKB-03 & 100 & 8.9 & 200& 8.2 & 290& 3.0\\
LKB-04 & 100 & 4.0 & 200& 3.0 & 290& $-1.0$\\ 
LKB-05 & 100 & $-2.0$ & 200&$-5.0$& 270& $-5.5$\\ 
\hline 
\end{tabular}

\caption{Table listing Galactic coordinates for the spline knots constructed to trace newly found structures.} 
\label{tab:hresult}
\end{table}

In Figure 3, we show how these two distance sequences cluster in angular momentum and energy space. The bright RC sequence (blue points) shows a wider spread in energy compared to the faint RC sequence, which has almost constant energy similar to phase-mixing structures identified in the solar neighbourhood \citep[e.g.][]{minchev09}. In order to make a possible link with recently uncovered kinematic ridges at intermediate distances in the Gaia eDR3 science verification study of the Anticenter \citep{ 2021arXiv210105811G}, we show how these two RC sequences look in the $V_{\phi}-R$ plane. This is shown in Figure 4, where we notice that the faint sequence (RC Seq 2) follows a continuation of the ``AC new ridge 2'' of \cite{ 2021arXiv210105811G}, whereas the brighter component shows larger spread perhaps belonging to an intervening hotter component of the disc being simultaneously picked up along the line of sight.

Similarly, for structures LKB-01 and LKB-02, we confirm that these are indeed similar to and associated with the TriAnd cloud \citep{rocha-pinto04} historically detected as a broad overdensity of M-giants in the direction $100<l<160$ and $-30<b<-15$. The association is confirmed through a detection of a clear RC sequences in the range $100<l<160$ with $16<G_{0}<17.5$ , corresponding to distances of $13<d/\,\rm{kpc}<25$ assuming $M_{G}=0.44$ for the RC \citep{hawkins17}, which are similar to estimates in \cite{rocha-pinto04} using M-giants. Moreover, the RC sequence shares similar line-of-sight velocities $-150<v_{los}/[\rm{km\,s^{-1}}]<-50$ as the RGBs in TriAnd in \citep{bergemann18} in the range $100<l<160$  with a median metallicity of $<[\rm{Fe/H}]>\sim-0.7$. Our detection suggests that TriAnd and its neighbouring Perseus overdensity in \cite{rocha-pinto04} are in fact all connected forming a long coherent structure in the disc (LKB-02) from $l\sim70$ to $l\sim250$. The disc nature of TriAnd was confirmed through M-giant/RR Lyrae ratio measurements in \cite{price-whelan15} and chemical tagging in \cite{bergemann18}.

\section{Interpretation: insights from numerical simulations}
Numerical N-body and cosmological simulations of interacting satellites with Milky-Way like hosts predict the formation of bending waves which propagate outwards and dissipate at the edge of the galaxy where the self-gravity is negligible which results in gradual heating but also corrugations/warping in the disc \citep[e.g.][]{gomez16,laporte18b}. These interactions also excite substructure within the disc, in the form of tidal tails and groups of stars undergoing coherent oscillations \citep[``feathers'', see Figures 1, 4 and 9 of][]{laporte19a}. Due to the long orbital timescales in the outer disc $T_{orb}\geq1\,\rm{Gyr}$, these remain coherent over several Gyrs ($t\sim6$\,Gyr). Using the ACS as a case study, \cite{laporte20} showed that this structure had a metallicity distribution and magnesium abundances consistent with the alpha-poor thin disc and a cumulative age distribution reminiscent of a truncated star formation history with predominantly old stars $t>10\,\rm{Gyr}$ when compared to stars in the more diffuse/extended Monoceros Ring structure. At face value this suggests that an extended thin disc might have already been in place at $z\sim2$ \citep{laporte20}.  Reproducing the physical extent of havoc in the outer disc requires invoking a larger progenitor halo masses for Sgr than historically assumed \citep[$M_{halo}\geq6\times10^{10}\,\rm{M_{\odot}}$, see][]{gibbons16, laporte18b} which is further supported by the recent discovery of a large number of globular clusters (GC) associated with Sgr \citep{minniti21} increasing the specific frequency of GCs, hence progenitor halo mass \citep[e.g.][]{hudson14}.

One possible interpretation would be that the structures uncovered here represent a new hierarchy of feathers in the disc, previously unknown due to the intervening dust. An alternative would be that not all the uncovered filamentary over-densities correspond to feathers, but may instead be projections of distant folds from the corrugated/flared outer disc \citep{xu15,price-whelan15, li17, thomas19}.

There are two possible complementary ways to discriminate between the two scenarios. One would be to measure the velocity dispersions for these structures from 6-D phase-space information and compare how kinematically cold they are with respect to their surroundings. Another method would be to measure photo-spectroscopic ages in these features in MSTOs and compare cumulative age distributions with the surrounding outer disc. Such a programme would be feasible with massively multiplexed facilities (e.g. WEAVE, DESI, PFS, MSE) and would provide important clues on the assembly of the Galactic thin disc and constraints for modelling of past satellite interactions \citep[e.g. Sgr or Antlia II,][]{laporte18b, vasiliev21, chakrabarti19}. 

\section{Conclusions}

Using Gaia eDR3 proper motions, we report the discovery of long thin disc overdensity structures with striking stream-like morphology in the outer disc permeating the Anticenter showing remarkable resemblance to predicted structures in N-body simulations of MW-like galaxies interacting with satellites \citep{laporte19a}. These bring a new and updated view on the intricate structure and dynamical origins of the outer disc in particular towards the so-called Monoceros Ring Complex which shows a succession of nested filamentary structures at different distances $d$ ranging from 10 to 30 kpc. While these are interpreted as tidally excited disc material with a narrow range of energies undergoing phase-mixing, we cannot certainly rule out the possibility that some may possibly be folds from a corrugating disc. Distinguishing between the these two hypotheses will require further modelling of the outer disc and observational efforts to mapping the full 6D phase-space information and ages (e.g. via main-sequence turn-off stars) in those structures and their surroundings. If these structures are created by interactions with external perturbers, they open an exciting possibility to age date impacts from satellites.  Upcoming spectroscopic surveys such as WEAVE, SDSS-V, DESI and PFS will  bring the outer disc into much better focus in the upcoming years allowing a more accurate characterisation of these newly uncovered structures with complementary radial velocity and abundance measurements offering an avenue to probe different stages in the interaction lifetime of the disc to satellite perturbers/perturbations.

\section*{Acknowledgements}
This work was started made use of data from the European Space Agency (ESA) mission {\it Gaia} (\url{https://www.cosmos.esa.int/gaia}), processed by the {\it Gaia} Data Processing and Analysis Consortium (DPAC, \url{https://www.cosmos.esa.int/web/gaia/dpac/consortium}). Funding for the DPAC has been provided by national institutions, in particular the institutions participating in the {\it Gaia} Multilateral Agreement. This paper made use of the Whole Sky Database (wsdb) created by S. Koposov. This work was supported in part by World Premier International Research Center Initiative (WPI Initiative), MEXT, Japan. The data underlying this article will be shared on reasonable request to the corresponding author.

\bibliographystyle{mnras}
\bibliography{master2.bib}{}
\bsp	
\label{lastpage}
\end{document}